\documentclass[twocolumn,showpacs,preprintnumbers,amsmath,amssymb]{revtex4}
\usepackage{epsfig}
\usepackage{graphicx}
\usepackage{dcolumn}
\usepackage{bm}
\usepackage{amsthm}
\usepackage{amsfonts}

\newcommand{\ket}[1]{\mbox{$ | #1 \rangle $}}
\newcommand{\bra}[1]{\mbox{$ \langle #1 | $}}

\begin{document}

\title{Upper bounds on the performance of differential-phase-shift
quantum key distribution}

\author{Hip\'olito G\'omez-Sousa and Marcos Curty} \affiliation{
ETSI Telecomunicaci\'on, 
University of Vigo, Campus Universitario, 36310 Vigo, Spain}

\date{\today}

\begin{abstract}
In this paper, we investigate limitations imposed by sequential attacks on the performance of 
a differential-phase-shift (DPS) quantum key distribution (QKD) protocol with weak coherent pulses. 
Specifically, we analyze a sequential attack based on optimal unambiguous discrimination of the relative 
phases between consecutive signal states emitted by the source.  We show that this attack can provide 
tighter upper bounds for the security of a DPS QKD scheme than those derived 
from sequential attacks where the eavesdropper aims to identify the state of each signal emitted by the source
unambiguously. 
\end{abstract}


\maketitle

\section{INTRODUCTION}

The main security threat of quantum key distribution (QKD) protocols based on weak coherent pulses (WCP) 
arises from the fact that some signals contain more than one photon prepared in the same polarization state. 
In this situation, the eavesdropper (Eve) can perform, for instance, the so-called {\it Photon Number Splitting} (PNS) 
attack on the 
multi-photon pulses \cite{Huttner95}. As a result, it turns out that the BB84 protocol \cite{BB84} with WCP can 
give a key generation rate of order $O(\eta^2)$, where $\eta$ denotes the transmission efficiency of the 
quantum channel \cite{inamori,inamori2}.

To obtain higher secure key rates over longer distances, different practical QKD schemes, that are robust against the 
PNS attack, have been proposed in recent years. One of these schemes is the so-called decoy-states 
\cite{decoy_t}, where the sender (Alice) randomly varies the mean photon number of the signal states 
that are forwarded to the receiver (Bob). This method can
deliver a secure key rate of order $O(\eta)$. Another possibility is based on the 
transmission of two non-orthogonal coherent states together with a strong reference pulse \cite{ben92}. 
This technique also provides a key generation rate of order $O(\eta)$ \cite{koashi04}. Finally, another 
potential approach is to use a differential-phase-shift (DPS) QKD protocol 
\cite{dpsqkd,dpsqkd2}. In
this scheme, Alice sends to Bob a train of WCP whose phases are randomly modulated by $0$ or $\pi$. On the 
receiving side, Bob measures out each incoming signal by means of an interferometer whose path-length 
difference is set equal to the time difference between two consecutive pulses. In this last case, however, a 
secure key rate of order $O(\eta)$ has only been proven so far against a special type of individual attacks 
where Eve acts and measures {\it photons} individually, rather than {\it signals}
\cite{dpsqkd2}, and also against a particular class of collective attacks where Eve attaches ancillary 
systems to each pulse or to each pair of successive pulses sent by Alice \cite{cyril_new}. While a complete security proof of 
a DPS QKD protocol against the most general attack is still missing, recently it has been shown that sequential attacks 
\cite{dpsqkd2} already impose strong restrictions on the performance of this QKD scheme with WCP. For instance, in \cite{curty_dps,curty_dps2,curty_dps3} 
it was proven that the 
long-distance implementations of DPS QKD reported in \cite{dpsqkd_exp1,dpsqkd_exp2,dpsqkd_exp2b,dpsqkd_exp3} 
would be insecure 
against a sequential attack based on unambiguous state discrimination (USD) of Alice's signal states 
\cite{usd,chef,jahma01}. 

In this paper, we investigate a novel sequential attack where Eve realizes unambiguous discrimination 
of the relative phases 
between Alice's signal states, and we obtain ultimate upper bounds on the maximal distance 
achievable by a DPS QKD scheme as a function of 
the error rate in the sifted key, and the mean photon number of the signals sent by Alice. It states that no
key distillation protocol can provide a secret key from the correlations established by the users. 
Moreover, we show that this attack can provide 
tighter upper bounds for the security of a DPS QKD scheme than those derived 
from a sequential attack where Eve performs USD of each signal state emitted by Alice 
\cite{curty_dps,curty_dps2,curty_dps3}.  

The paper is organized as follows. In section~\ref{sec_1} we describe in more detail a DPS QKD protocol. 
Then, in section~\ref{seqattack}, we present a sequential attack against this QKD scheme 
based on optimal unambiguous discrimination of the relative phases between Alice's signal states. Here 
we obtain upper bounds on the performance of a DPS QKD scheme as a function 
of the error rate in the sifted key and 
the mean photon number of Alice's signal states. Finally, section~\ref{CONC} concludes the paper with a 
summary. The manuscript contains as well one appendix with additional calculations.

\section{Differential-phase-shift QKD}\label{sec_1}

\begin{figure}
\begin{center}
\includegraphics[scale=1.1]{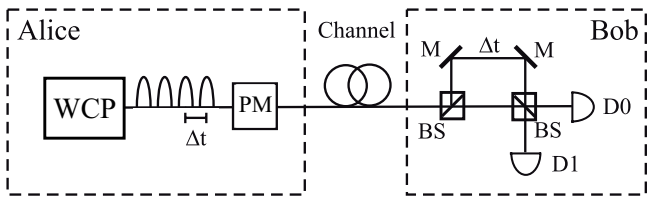}
\end{center}
\caption{Basic setup of a DPS QKD scheme. PM denotes a phase modulator, BS, a $50:50$ beam splitter, M, 
a mirror, D0 and D1 are two photon detectors and $\Delta{}t$ represents the time difference between two consecutive 
pulses. \label{dpsqkd}}
\end{figure}

The basic setup is illustrated in figure~\ref{dpsqkd}.
Alice prepares first a train of coherent states $\ket{\alpha}$ and, afterwards, she modulates, at random and independently 
every time, the phase of each pulse to be $0$ or $\pi$. As a result, she produces a random train of signal states $\ket{\alpha}$ 
or $\ket{-\alpha}$ that are sent to Bob through the quantum channel. On the receiving side, Bob uses a $50:50$ beam splitter 
to divide the incoming pulses into two possible paths and then he recombines them again using another $50:50$ beam 
splitter. The time delay introduced by Bob's interferometer is set equal to the time difference $\Delta{}t$ between two 
consecutive pulses. 
Whenever the relative phase between two consecutive signals is $0$ ($\pm\pi$) only the photon detector $D0$ ($D1$) may 
produce a ``click" (at least one photon is detected). For each detected event, Bob records the time slot where he obtained a 
click and the actual detector that fired.

Once the quantum communication phase of the protocol is completed, Bob uses a classical authenticated channel 
to announce the time 
slots where he obtained a click, but he does not reveal which detector fired each time. From this information provided by Bob,
together with the knowledge of the phase value used to modulate each pulse, Alice can infer which photon detector had 
clicked at Bob's side each given time. Then, Alice and Bob agree, for instance, to select a bit value ``0" whenever the 
photon detector $D0$ fired, and a bit value ``1" if the detector $D1$ clicked. In an ideal scenario, Alice and Bob end up 
with an identical string of bits representing the sifted key. Due to the noise introduced by the quantum channel, together 
with possible imperfections of Alice and Bob's devices, however, the sifted key typically contains some errors. Then, Alice 
and Bob perform error-correction to reconcile the data and privacy amplification to decouple the data from Eve. (See, for 
instance, \cite{gisin_rev_mod}.)

\section{Sequential attacks against differential-phase-shift QKD}
\label{seqattack}

A sequential attack can be seen as a special type of intercept-resend attack \cite{dpsqkd2,curty_dps,curty_dps2,curty_dps3}. 
First, Eve measures out every signal state emitted by Alice with a detection apparatus located very close to the sender. 
Afterwards, she 
transmits each measurement result through a lossless classical channel to a source close to Bob. Whenever Eve
obtains a predetermined number of consecutive {\it successful} measurement outcomes, this source prepares a new 
train of non-vacuum signal states that is forwarded to Bob. 
Otherwise, Eve typically sends vacuum signals to Bob to avoid errors \cite{note}. Whether a measurement result is considered to be successful or not, and which type of signal 
states Eve sends to Bob, depends on Eve's particular eavesdropping strategy and on her measurement device. 
Sequential attacks transform 
the original quantum channel between Alice and Bob into an entanglement breaking channel \cite{Horodecki03} and, 
therefore, they do not allow the distribution of quantum correlations needed to establish a secret key \cite{Curty04}.

The first sequential attack against a DPS QKD protocol was introduced very briefly in \cite{dpsqkd2}. In this proposal, 
Eve employs a 
detection apparatus equivalent to Bob's setup. A successful result is associated with Eve obtaining a click in her 
measurement device. This click identifies unambiguously the relative phase ($0$ or $\pm\pi$) between two consecutive 
pulses emitted by Alice and, therefore, it reveals Eve the bit value encoded by the sender. A failure corresponds 
to the absence of a click. 
However, since Alice emits WCP with typical average photon number quite low, so is the 
probability that Eve obtains a successful result in this scenario. In order to increase Eve's successful probability other
sequential attacks have been proposed more recently \cite{curty_dps,curty_dps2,curty_dps3}. 
These attacks are typically based on Eve 
realizing USD of each signal state emitted by Alice, since Eve can always access a local oscillator that is
phase-locked to the coherent light source employed by the sender \cite{curty_dps3}. 
In particular, when Eve identifies unambiguously a signal state emitted by 
Alice, {\it i.e.}, she determines without error whether it is $\ket{\alpha}$ or $\ket{-\alpha}$, 
then she considers this result as successful. Otherwise, she considers it a failure. In \cite{curty_dps}
it was shown that this class of sequential attacks can provide tighter 
upper bounds on the performance of a DPS QKD protocol than 
those derived from a sequential attack where Eve uses the same measurement apparatus like Bob. 

In this section, we introduce an improved version of the sequential attack proposed in \cite{dpsqkd2}, and we
investigate again the situation  
where Eve tries to identify the relative phases between Alice's signal states unambiguously. 
As a result, it turns out that the attack we propose can provide stronger limitations for the security of a DPS QKD 
scheme than those reported in \cite{curty_dps,curty_dps2,curty_dps3}. 
In our analysis we consider a conservative definition of 
security, {\it i.e.}, we assume that Eve can always control some flaws in Alice's and Bob's devices ({\it e.g.}, the 
detection efficiency, the dark count probability and the dead-time of the detectors), together with the losses in 
the channel, and she exploits them to obtain maximal information about the shared key.

\subsection{Optimal unambiguous discrimination between relative phases}\label{meas}

In a DPS QKD protocol Alice sends to Bob a train of WCP each of them 
prepared in the state $\ket{\alpha}$ or  $\ket{-\alpha}$. These coherent states can be expressed in some orthogonal 
basis $\{\ket{0}, \ket{1}\}$ as follows
\begin{equation}
\ket{\pm\alpha}=a\ket{0}\pm{}b\ket{1},
\end{equation}
where we assume, without loss of generality, that the coefficients $a$ and $b$ are given by
\begin{eqnarray}\label{coef}
a&=&\sqrt{\frac{1}{2}[1+\exp{(-2\mu_{\alpha})}]}, \nonumber \\
b&=&\sqrt{\frac{1}{2}[1-\exp{(-2\mu_{\alpha})}]}, 
\end{eqnarray}
with $\mu_{\alpha}=\vert\alpha\vert^2$ denoting the mean photon number of Alice's signal states. That is, 
$a$ and $b$ satisfy: $a\in\mathbb{R}$, $b\in\mathbb{R}$, $a^2+b^2=1$, and $a>b$ when $\mu_\alpha\neq{}0$.

The state of a block of $M$ consecutive WCP emitted by Alice, that we shall denote as 
$\ket{\psi(\vec x_M)}$, can be written as
\begin{eqnarray}\label{uno}
\ket{\psi(\vec x_M)}&=&\bigotimes_{i=1}^M\ket{(-1)^{x_i}\alpha}=\sum_{n_1, ..., n_M=0}^1 
(-1)^{\sum_{i=1}^M x_in_i} \nonumber \\
&\times&{}a^{M-\sum_{i=1}^M n_i} b^{\sum_{i=1}^M n_i} \ket{n_1, ..., n_M},
\end{eqnarray}
with the coefficients $a$ and $b$ given by (\ref{coef}), and
where the vector $\vec x_M=(x_1, ..., x_M)$, with $x_i\in\{0,1\}$, contains 
the information about the value of the phase ($0$ or $\pi$) imprinted by Alice in each pulse within the block. 

In order to access to the relative phase information encoded in a block of signals sent by Alice, however, 
it is not necessary to completely identify the vector $\vec x_M$. For instance, the relative phase between 
pulse $N$ and 
pulse $N-1$  in $\ket{\psi(\vec x_M)}$, with $2\leq{}N\leq{}M$, is simply given by $0$ ($\pm\pi$) when 
$x_{N}\oplus{}x_{N-1}=0$ ($1$). In general, for any given state $\ket{\psi(\vec x_M)}$, there exists always 
another state $\ket{\psi(\vec x_M\oplus\vec 1_M)}$, with $\vec x_M\oplus\vec 1_M=(x_1\oplus{}1, ..., x_M\oplus{}1)$, 
that has 
precisely the same $M-1$ relative phases as $\ket{\psi(\vec x_M)}$. This means, in particular, that the problem 
of determining the relative 
phases of Alice's signal states can be formulated as a discrimination problem between $2^{M-1}$ 
mixed states given by
\begin{eqnarray}\label{std}
\rho(\vec x_M)&=&\frac{1}{2}\bigg(\ket{\psi(\vec x_M)}\bra{\psi(\vec x_M)}\nonumber \\
&+&\ket{\psi(\vec x_M\oplus\vec 1_M)}\bra{\psi(\vec x_M\oplus\vec 1_M)}\bigg),
\end{eqnarray}
with the coefficient $x_M=0$. That is, the vector $\vec x_M$ has now the form  
\begin{equation}\label{vecx}
\vec x_M=(x_1, ..., x_{M-1},0), 
\end{equation}
with $x_i\in\{0,1\}$. This last condition arises because $\rho(\vec x_M)$ satisfies
$\rho(x_1, ..., x_{M-1},0)=\rho(x_1\oplus{}1, ..., x_{M-1}\oplus{}1,1)$. The normalization term $\frac{1}{2}$ that appears 
in (\ref{std}) is due to the fact that all the states $\ket{\psi(\vec x_M)}$ have equal a priori probabilities.

To distinguish between the signals states given by (\ref{std}), we shall consider that Eve follows a USD strategy. 
That is, the constraint is that the 
measurement employed by Eve should never wrongly identify a state $\rho(\vec x_M)$, but it can provide sometimes an 
inconclusive result \cite{usd,chef,jahma01}. The goal is to keep the fraction of inconclusive outcomes as low 
as possible.  

Let the set of binary vectors $V_{y,M}$, with $y\in\{A,B\}$, be defined as 
$V_{y,M}=\{(n_1, ..., n_M)\mid{}n_i\in\{0, 1\},\textrm{ and } \sum_{i=1}^M n_i 
\textrm{ even if $y=A$, odd if $y=B$}\}$,  
and let $\mathcal{Y_M}$ denote the subspace spanned by the orthogonal 
states $\{\ket{n_1, ..., n_M}\}$, with the vectors $(n_1, ..., n_M)\in{}V_{y,M}$. The signal states $\rho(\vec x_M)$ 
given by (\ref{std}) can be written in a block-diagonal form as
\begin{equation}\label{yns}
\rho(\vec x_M)=\sum_{y\in\{A,B\}} p_{y,M}\ket{\psi_{y}(\vec x_M)}\bra{\psi_{y}(\vec x_M)},
\end{equation}
where the probabilities $p_{y,M}$ are given by
\begin{equation}\label{prob}
p_{y,M}=\sum_{\substack{n_1, ..., n_M=0\\\vec n_M\in{}V_{y,M}}}^1
\bigg(a^{M-\sum_{i=1}^M n_i} b^{\sum_{i=1}^M n_i}\bigg)^2,
\end{equation}
with the vector $\vec n_M\equiv{}(n_1, ..., n_M)$, and where the states $\ket{\psi_{y}(\vec x_M)}$ have the form
\begin{eqnarray}\label{states_ja}
\ket{\psi_{y}(\vec x_M)}&=&\frac{1}{\sqrt{p_{y,M}}}
\sum_{\substack{n_1, ..., n_M=0\\\vec n_M\in{}V_{y,M}}}^1
(-1)^{\sum_{i=1}^{M-1} x_in_i}\nonumber \\
&\times&a^{M-\sum_{i=1}^M n_i}b^{\sum_{i=1}^M n_i}
\ket{n_1, ..., n_M}.
\end{eqnarray}
That is, the signals $\ket{\psi_{y}(\vec x_M)}\in\mathcal{Y_M}$. 

This means, in particular, that we can always assume, without loss of generality, that Eve's measurement strategy 
includes an initial step which projects the mixed states $\rho(\vec x_M)$ onto the orthogonal subspaces  
$\mathcal{A_M}$ and $\mathcal{B_M}$. This projective measurement is characterized by the following two operators:
\begin{equation}
\Pi_{y,M}=
\sum_{\substack{n_1, ..., n_M=0\\\vec n_M\in{}V_{y,M}}}^1
\ket{n_1, ..., n_M}\bra{n_1, ..., n_M},
\end{equation}
with $y\in\{A,B\}$. It satisfies the condition $[\textrm{Tr}{(\Pi_{y,M}\rho(\vec x_M))}]^{-1}\Pi_{y,M}\rho(\vec x_M)\Pi_{y,M}^\dag=
\ket{\psi_{y}(\vec x_M)}\bra{\psi_{y}(\vec x_M)}$. That is, it outputs the state $\ket{\psi_{y}(\vec x_M)}$
with probability $p_{y,M}$. 

The question of discriminating the $2^{M-1}$ mixed states given by (\ref{std}) can then be
reduced to the problem of distinguishing $2^{M-1}$ pure states $\ket{\psi_{y}(\vec x_M)}$.
To discriminate between the signals $\ket{\psi_{y}(\vec x_M)}$, we shall consider a measurement strategy which can
involve at most $M-1$ steps. Before providing the exact details of the measurement, let us sketch very briefly 
its principal parts. Eve starts by performing a filter operation on $\ket{\psi_{y}(\vec x_M)}$. If the filter operation 
succeeds, Eve 
obtains $x_{M-1}\oplus{}x_{M}$. That is, Eve learns with certainty the relative phase between the first two pulses in 
the block. Moreover, this filter operation also outputs a quantum state which still contains complete information 
about the remaining 
$M-2$ relative phases within the block. On the contrary, if the filter operation fails, the value of
$x_{M-1}$ is not accessible anymore, and Eve cannot obtain the first two relative phases ({\it i.e.}, 
$x_{M-1}\oplus{}x_{M}$, 
and $x_{M-2}\oplus{}x_{M-1}$) within the block. In this last case, however, the filter operation outputs
a state which contains information about the remaining $M-3$ relative phases within the block. Eve repeats 
the same procedure several times, but now applied to the quantum state provided by the filter operation in the previous step. 
To gain full information about {\it all} the relative phases contained in $\ket{\psi_{y,M}(\vec x)}$, Eve needs to 
obtain $M-1$ consecutive successful filtering results. 

The main motivation to select such a particular implementation 
of a USD measurement is closely related to Eve's eavesdropping strategy, which will be introduced in 
section \ref{attack}. The principal idea behind this method is that, with some finite probability, Eve can always determine 
the value of some relative phases in $\ket{\psi_{y}(\vec x_M)}$, even if she is not able 
to identify all of them. Moreover, as we show in \ref{ap_opt}, it turns out that this measurement strategy is optimal, {\it i.e.},
it minimizes the probability of obtaining an inconclusive result when distinguishing all the $M-1$ relative phases 
of Alice's signal states. Next, we provide the technical details of Eve's measurement.  

The set of $M-1$ possible filter operations employed by Eve is defined by the following two Kraus operators:
\begin{eqnarray}\label{filter}
F_{succ,y,N}&=&G_{y,N-1}\otimes\ket{0}\bra{0}+I_{N-1}\otimes\ket{1}\bra{1}, \\
F_{fail,y,N}&=&(I_{N-1}-G_{y,N-1}^\dag{}G_{y,N-1})^{1/2}\otimes\ket{0}\bra{0}, \nonumber 
\end{eqnarray}
with $2\leq{}N\leq{}M$, and where $I_{N-1}$ denotes the identity operator in $\mathcal{H}_{2^{N-1}}$, and 
the operator $G_{y,N-1}$ is given by
\begin{eqnarray}
{}G_{y,N-1}&=&
\sum_{\substack{n_1, ..., n_{N-1}=0\\\vec n_{N-1}\in{}V_{y,N-1}}}^1
\bigg(\frac{b}{a}\bigg)^{2(n_{N-1}\oplus{}1)}\\
&\times&
\ket{n_1, n_2, ..., n_{N-1}\oplus{}1}\bra{n_1, n_2, ..., n_{N-1}}.\nonumber
\end{eqnarray}

Let $\ket{\phi_{y}(\vec x_N)}$ denote a quantum state of the form
\begin{eqnarray}\label{states_tilde}
\ket{\phi_{y}(\vec x_N)}&=&\frac{1}{\sqrt{p_{y,N}}}
\sum_{\substack{n_1, ..., n_{N}=0\\\vec n_{N}\in{}V_{y,N}}}^1
(-1)^{\sum_{i=1}^{N-1} (x_i\oplus{}x_ N)n_i}\nonumber \\
&\times&a^{N-\sum_{i=1}^N n_i}b^{\sum_{i=1}^N n_i}
\ket{n_1, ..., n_N},
\end{eqnarray}
with $1\leq{}N\leq{}M$. That is, when $N=M$ these states satisfy $\ket{\phi_{y}(\vec x_M)}=\ket{\psi_{y}(\vec x_M)}$ for all 
$\vec x_M$ given by (\ref{vecx}). Let $\vec x_{N-1}$ denote the vector that is formed by the first $N-1$ elements of 
$\vec x_M$.  For any $N$ satisfying $2\leq{}N\leq{}M$, the signal states given by (\ref{states_tilde}) 
can be written as a function of $\ket{\phi_{y}(\vec x_{N-1})}$ and $\ket{\phi_{{\bar y}}(\vec x_{N-1})}$, 
with $\bar y=B$ when $y=A$ and $\bar y=A$ when $y=B$, as
\begin{eqnarray}\label{cans}
\ket{\phi_{y}(\vec x_N)}&=&\frac{1}{\sqrt{p_{y,N}}}\Big(
a\sqrt{p_{y,N-1}}\ket{\phi_{y}(\vec x_{N-1})}\ket{0}\\
&+&(-1)^{x_{N-1}\oplus{}x_{N}}b\sqrt{p_{{\bar y},N-1}}\ket{\phi_{{\bar y}}(\vec x_{N-1})}\ket{1}\Big), \nonumber
\end{eqnarray}
up to a global phase.

Suppose now that the filter operation defined by (\ref{filter}) receives as input the state 
$\ket{\psi_{y}(\vec x_M)}\equiv{}\ket{\phi_{y}(\vec x_M)}$. The probability 
of getting a successful result, that we shall represent as $p_{succ,y,M}$, can be calculated as
$p_{succ,y,M}=\bra{\phi_{y}(\vec x_M)}F_{succ,y,M}^\dag{}F_{succ,y,M}\ket{\phi_{y}(\vec x_M)}$. This quantity 
is given by
$p_{succ,y,M}=(p_{y,M})^{-1}2b^2p_{{\bar y},M-1}$. 
If the filter operation succeeded, the resulting normalized filtered state, that we shall denote as 
$\ket{\phi_{succ,y}(\vec x_M)}$, can be calculated as
$\ket{\phi_{succ,y}(\vec x_M)}=(\sqrt{p_{succ,y,M}})^{-1}F_{succ,y,M}\ket{\phi_{y}(\vec x_M)}$. We obtain 
$\ket{\phi_{succ,y}(\vec x_M)}=\ket{\phi_{{\bar y}}(\vec x_{M-1})}\otimes\ket{\psi_{y,M}}$,
with the state $\ket{\psi_{y,M}}$ given by $\ket{\psi_{y,M}}=(\sqrt{2})^{-1}[\ket{0}+(-1)^{x_{M-1}\oplus{}x_M}\ket{1}]$,
up to a global phase. That is, the relative phase between 
pulse $M$ and pulse $M-1$ is now completely accessible to Eve. She only has to measure the state 
$\ket{\psi_{y,M}}$ in 
the orthogonal basis $\ket{\pm}=(\sqrt{2})^{-1}(\ket{0}\pm{}\ket{1})$ to learn its value.     

On the contrary, the probability of obtaining a failure, that we shall denote as $p_{fail,y,M}$, can be calculated as
 $p_{fail,y,M}=\bra{\phi_{y}(\vec x_M)}F_{fail,y,M}^\dag{}F_{fail,y,M}\ket{\phi_{y}(\vec x_M)}$. 
This quantity is given by $p_{fail,y,M}=(p_{y,M})^{-1}(1-2b^2)p_{y,M-2}=1-p_{succ,y,M}$.
Whenever the filter operation failed, the resulting normalized filtered 
state, that we shall denote as $\ket{\phi_{fail,y}(\vec x_M)}$, can be calculated as
 $\ket{\phi_{fail,y}(\vec x_M)}=(\sqrt{p_{fail,y,M}})^{-1}F_{fail,y,M}\ket{\phi_{y}(\vec x_M)}$. 
We obtain $\ket{\phi_{fail,y}(\vec x_M)}=\ket{\phi_{y}(\vec x_{M-2})}\otimes\ket{00}$,
up to a global phase. That is, if Eve fails when filtering the state 
$\ket{\phi_{y}(\vec x_M)}$, then the value of $x_{M-1}$ is not accessible to her anymore, and Eve cannot 
obtain the relative phase information between pulse $M$ and pulse $M-1$, and also between pulse 
$M-1$ and pulse $M-2$, within the block.

Once the first filter operation finished, Eve is left with a quantum state which contains the signal  
$\ket{\phi_{{\bar y}}(\vec x_{M-1})}$ if the filter succeeded, or the signal $\ket{\phi_{y}(\vec x_{M-2})}$ if it failed.
Then, she can repeat the same procedure again, and filter these signal states 
to try to obtain $x_{M-2}\oplus{}x_{M-1}$ if the original state was $\ket{\phi_{{\bar y}}(\vec x_{M-1})}$, 
or $x_{M-3}\oplus{}x_{M-2}$ if it was $\ket{\phi_{y}(\vec x_{M-2})}$. In general, 
whenever a filter operation given by (\ref{filter}) receives as input the state 
$\ket{\phi_{y}(\vec x_N)}$, with $2\leq{}N\leq{}M$, then the probability of getting a successful result is given by
\begin{equation}\label{lms}
p_{succ,y,N}=2b^2\frac{p_{{\bar y},N-1}}{p_{y,N}}.
\end{equation}
If the filter operation succeeded, the resulting normalized filtered state has the form
\begin{equation}
\ket{\phi_{succ,y}(\vec x_N)}=\ket{\phi_{{\bar y}}(\vec x_{N-1})}\otimes\ket{\psi_{y,N}},
\end{equation}
with the signal $\ket{\psi_{y,N}}$ given by
\begin{equation}
\ket{\psi_{y,N}}=\frac{1}{\sqrt{2}}\Big[\ket{0}+(-1)^{x_{N-1}\oplus{}x_N}\ket{1}\Big], 
\end{equation}
up to a global phase. On the contrary, the probability of obtaining a failure can be expressed as
\begin{equation}\label{pfail}
p_{fail,y,N}=(1-2b^2)\frac{p_{y,N-2}}{p_{y,N}},
\end{equation}  
with the probabilities $p_{A,0}\equiv{}1$ and $p_{B,0}\equiv{}0$. In this last case, the resulting normalized filtered 
state is given by
\begin{equation}\label{failure}
\ket{\phi_{fail,y}(\vec x_N)}=\ket{\phi_{y}(\vec x_{N-2})}\otimes\ket{00},
\end{equation}
up to a global phase. 

Let us now calculate the probability that Eve learns the first $k\in[1,M-1]$ relative phases of $\rho(\vec x_M)$.
As we have seen above, to obtain the relative phase between pulse $N$ and pulse $N-1$ within a block
of $M$ signals sent by Alice, Eve has to successfully filter a state of the form $\ket{\phi_{y}(\vec x_N)}$. 
Let $p_{succ,N}$ denote the probability that Eve obtains the value of $x_{N-1}\oplus{}x_N$ conditioned on the fact
that Eve has access to a signal $\ket{\phi_{y}(\vec x_N)}$, with $y\in\{A,B\}$. This probability can be written as
\begin{equation}\label{ssm}
p_{succ,N}=\sum_{y\in\{A,B\}}p_{N}^yp_{succ,y,N},
\end{equation}
where $p_{N}^y$ represents the probability that the state filtered by Eve when trying to obtain $x_{N-1}\oplus{}x_N$ 
belongs to the subspace $\mathcal{Y_N}$.
When $N=M$, we have that $p_{M}^y$ is simply given by $p_{M}^y=p_{y,M}$, with $p_{y,M}$ of 
the form (\ref{prob}). This means, in particular, that $p_{succ,M}=2b^2(p_{y,M-1}+p_{{\bar y},M-1})=2b^2$, 
since $p_{y,M-1}+p_{{\bar y},M-1}=1$. If $N=M-1$, the probabilities $p_{M-1}^y$ can be expressed as 
$p_{M-1}^y=(p_{succ,M})^{-1}p_{M}^{\bar y}p_{succ,{\bar y},M}$. Using (\ref{lms}), together with the fact that 
$p_{succ,M}=2b^2$ and $p_{M}^{\bar y}=p_{{\bar y},M}$, we obtain $p_{M-1}^y=p_{y,M-1}$. That is, 
$p_{succ,M-1}$ is given by $p_{succ,M-1}=2b^2(p_{y,M-2}+p_{{\bar y},M-2})=2b^2$. Similarly, 
when $2\leq{}N\leq{}M-2$, the state $\ket{\phi_{y}(\vec x_N)}$ can only arise from a filter operation on a 
signal  $\ket{\phi_{\bar y}(\vec x_{N+1})}$ that succeeded, or from a filter operation on a signal 
$\ket{\phi_{y}(\vec x_{N+2})}$ that failed. If it comes from a successful filter operation on 
$\ket{\phi_{\bar y}(\vec x_{N+1})}$, then $p_{N}^y$ can be written as 
$p_{N}^y=(p_{succ,N+1})^{-1}p_{N+1}^{\bar y}p_{succ,{\bar y},N+1}$. Starting with the case $N=M-2$, we already 
showed that $p_{succ,M-1}=2b^2$ and $p_{M-1}^{\bar y}=p_{{\bar y},M-1}$. This means, therefore, that 
$p_{M-2}^y=p_{y,M-2}$. If the state $\ket{\phi_{y}(\vec x_N)}$ arises from a filter operation on 
$\ket{\phi_{y}(\vec x_{N+2})}$ which failed, then $p_{N}^y$ is given by 
$p_{N}^y=(p_{fail,N+2})^{-1}p_{N+2}^{y}p_{fail,y,N+2}$. Starting again with the case $N=M-2$, and using 
(\ref{pfail}) together with the fact that $p_{fail,M}=1-p_{succ,M}=1-2b^2$ and $p_{M}^{y}=p_{y,M}$, we have that $p_{M-2}^y=p_{y,M-2}$ also in this scenario. Finally, from (\ref{ssm}) we obtain that $p_{succ,M-2}$ satisfies 
$p_{succ,M-2}=2b^2$. Following a recursive argumentation, it is straightforward to show that 
\begin{equation}\label{psucc} 
p_{succ,N}=2b^2=1-\exp{(-2\mu_{\alpha})},
\end{equation}
for all $N$ satisfying $2\leq{}N\leq{}M$, and where in the last equality we have used (\ref{coef}).
This means, in particular, that the probability that Eve learns the first $k\in[1,M-1]$ relative phases of $\rho(\vec x_M)$
can now be expressed as
\begin{equation}
\prod_{i=0}^{k-1} p_{succ,M-i}=[1-\exp{(-2\mu_{\alpha})}]^{k},
\end{equation}
As already mentioned before, it can be proven that this measurement is optimal,  
{\it i.e.}, it minimizes the probability of having an inconclusive result when distinguishing all the relative phases 
of Alice's signal states. (See Appendix \ref{ap_opt}.)  

\subsection{Eavesdropping strategy}\label{attack}

For simplicity, we shall consider that Eve treats all the signal states sent by Alice as a single block of signals, and she 
tries to discriminate each relative phase within the block. Whenever she identifies unambiguously a 
predetermined number of consecutive relative phases sent by Alice, {\it i.e.},
she determines without error whether each relative phase is $0$ or $\pm\pi$, she considers this sequence of
measurement outcomes successful. Otherwise she considers it a failure. We define the integer parameter 
$M_{\rm min}$ as the minimum number of consecutive relative phases that Eve needs to correctly identify in order 
to consider the sequence of measurement outcomes successful. More precisely, if $k\geq{}0$ denotes the
total number of consecutive relative phases unambiguously identified by Eve before she obtains an inconclusive 
result, then, whenever $k>M_{\rm min}$, Eve prepares a new train of signal states that is forwarded to Bob. 
On the other hand, if $k<M_{\rm min}$ Eve sends to Bob $k+2$ vacuum states, where the last vacuum state 
corresponds to Eve's inconclusive result. Finally, whenever $k=M_{\rm min}$ we shall consider that Eve
employs a probabilistic strategy that combines the two previous ones. In particular, we assume that Eve sends 
to Bob a new train of signal states with probability $q$ and, with probability $1-q$, she sends to Bob
$M_{\rm min}+2$ vacuum states. That is, the parameter $q$ allows Eve to smoothly fit her eavesdropping 
strategy to the observed data \cite{curty_dps}. 

Moreover, for simplicity, 
we define the integer parameter $M_{\rm max}>M_{\rm min}$ as the maximum number of consecutive 
unambiguous discrimination successful results that Eve can obtain in order to send to Bob a train of signal 
states. That is, whenever Eve determines unambiguously $M_{\rm max}$ consecutive relative phases within a
block of them
then she discards the next two phases, sends to Bob a train of signal states, and begins again the measurement 
process of the remaining phases. The reason to discard two consecutive relative phases in this scenario is just to
guarantee that between any two blocks of signal states sent by Eve there always exists, at least,
one vacuum state.  Specifically, suppose, for instance, that after $M_{\rm max}$
successful results, Eve's filter operation outputs, with probability $p_{N}^y$, a state $\ket{\phi_{y}(\vec x_N)}$ 
given by (\ref{states_tilde}).
For $N>2$, the state $\ket{\phi_{y}(\vec x_N)}$ can be written as
\begin{eqnarray}
\ket{\phi_{y}(\vec x_N)}&=&\frac{1}{\sqrt{p_{y,N}}}\Big\{
\sqrt{p_{y,N-2}}\ket{\phi_{y}(\vec x_{N-2})}\big[a^2\ket{00}_C \nonumber \\
&+&(-1)^{x_{N-1}\oplus{}x_{N}}b^2\ket{11}_C\big]+ab\sqrt{p_{{\bar y},N-2}}\nonumber \\
&\times&\ket{\phi_{\bar y}(\vec x_{N-2})}\big[(-1)^{x_{N-2}\oplus{}x_{N}}\ket{01}_C\nonumber \\
&+&(-1)^{x_{N-2}\oplus{}x_{N-1}}\ket{10}_C\big]\Big\}, 
\end{eqnarray}
up to a global phase. If now Eve discards subsystem $C$, the resulting signal state can be expressed as
$\sum_{y\in\{A,B\}} p_{N}^y\textrm{Tr}_C{(\ket{\phi_{y}(\vec x_N)}\bra{\phi_{y}(\vec x_N)})}$. After some calculations, 
and using the fact that $p_{N}^y=p_{y,N}$ (see Section \ref{meas}), 
we obtain that this state is of the form given by
(\ref{yns}), with $M=N-2$. That is, the value of $x_{N-1}$ is not accessible anymore, but 
Eve can start again her measurement strategy on $\rho(\vec x_{N-2})$.

Let us now introduce the type of signal states that Eve forwards to Bob when she obtains 
$M_{\rm min}\leq{}k\leq{}M_{\rm max}$ consecutive successful measurement outcomes. 
To guarantee that Eve's presence remains unnoticeable to the legitimate users, she needs 
to select these signal states such that they can reproduce the statistics expected 
by the legitimate users after their measurements. 
For this, we shall consider the standard version of a DPS QKD protocol, where Alice and Bob only monitor 
the raw bit rate (before the key distillation phase) together with the time instances in which Bob
obtains a click. It was shown in \cite{curty_dps3} that the main limitation on the class of 
signal states that Eve can send to Bob in this scenario arises from the dead-time of BobÕs detectors. In 
particular, to be able to mimic the expected dead-time of the detectors, Eve 
has to select trains of signal states that can produce only one click on
Bob's side within a dead-time period \cite{note2}.
To achieve this goal, we shall assume that whenever Eve identifies $k$ consecutive relative phases encoded by Alice
then she chooses her signal states, that we denote as 
$\ket{ \psi_{\rm e}^{k}}$, containing only one photon distributed 
among $k+1$ temporal modes. These modes correspond to $k+1$ consecutive pulses sent by Alice, {\it i.e.}, 
the time difference between any two consecutive temporal modes is set equal to the time difference $\Delta{}t$
between two consecutive pulses. 
Specifically, we shall consider that the states $\ket{ \psi_{\rm e}^{k}}$ are given by \cite{curty_dps2,curty_dps3}
\begin{equation}\label{signal_uds}
\ket{\psi_{\rm e}^k}=\sum_{n=1}^{k+1} A_n^{(k)} \exp{(i\theta_n)}
\hat{a}_n^\dag\ket{vac},
\end{equation}
with the coefficients $A_n^{(k)}\in{}\mathbb{C}$ and where the normalization condition $\sum_{n=1}^{k+1}
\vert{}A_n^{(k)}\vert{}^2=1$ is always satisfied. The angles  $\theta_n$ are selected such that they reproduce
the relative phases identified by Eve's measurement, {\it i.e.}, $\theta_n-\theta_{n-1}$, with $1<n\leq{}k+1$, is equal 
to the relative phase between pulse $n$ and pulse $n-1$ sent by Alice. The operator $\hat{a}_n^\dag$ represents
a creation operator for one photon in temporal mode $n$, and the state $\ket{vac}$ refers to the vacuum state. 
The superscript $k$ labeling the coefficients $A_n^{(k)}$ emphasizes the fact that the value of these coefficients 
may depend on the number of temporal modes contained in $\ket{\psi_{\rm e}^k}$. 

Eve also appends some vacuum states to each signal $\ket{ \psi_{\rm e}^{k}}$. The main idea behind this 
procedure is to guarantee that whenever Bob obtains a click on his detection apparatus, then he cannot 
obtain any other click afterwards during a period of time at least equal to the dead-time of his detectors.
The minimum number of vacuum states that Eve needs to send to Bob after each signal 
$\ket{\psi_{\rm e}^k}$ is given by $1+d$, with $d=\lceil{}t_{\rm d}f_{\rm c}\rceil$, and where $t_{\rm d}$ and 
$f_{\rm c}$ denote, respectively, the dead-time of Bob's detectors and the clock frequency of the system 
\cite{curty_dps3}. The minimum value of $d$ arises from the case where Bob obtains a click in the last possible 
temporal mode. Whenever Eve forwards to Bob a state $\ket{ \psi_{\rm e}^{k}}$ together with 
$1+d$ vacuum states then she also has to discard some extra relative phases of $\ket{\phi_{y}(\vec x_N)}$
according to the procedure explained above before she begins again with her measurement of the 
remaining relative phases within the block.

\begin{figure}
\begin{center}
\includegraphics[scale=0.8]{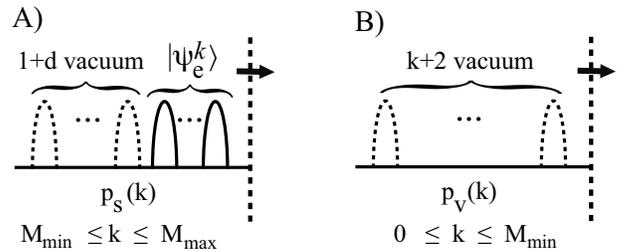}
\end{center}
\caption{Possible signal states that Eve sends to Bob together with their a priori probabilities. The arrow indicates the 
transmission direction.\label{strategy1_meas}}
\end{figure}

In section \ref{meas} we showed that, given $\rho(\vec x_M)$, the probability that Eve learns the first 
$k\in[1,M-1]$ relative phases of $\rho(\vec x_M)$ is given by $p^{k}$ with 
\begin{equation}\label{spusd2x}
p=1-\exp{(-2\mu_{\alpha})}. 
\end{equation}
This means, in particular, that the probability that Eve sends to Bob a train of signal states $\ket{ \psi_{\rm e}^{k}}$,
together with $1+d$ vacuum states, is given by
\begin{equation}\label{ps_k}
p_{\rm s}(k) = \left\{ \begin{array}{ll} q p^{M_{\rm min}}(1-p) &
\textrm{if $k=M_{\rm min}$}\\
p^k(1-p) & \textrm{if $M_{\rm min}<k<M_{\rm max}$}\\
p^{M_{\rm max}} & \textrm{if $k=M_{\rm max}$}\\
0 & \textrm{otherwise,}
\end{array} \right.
\end{equation}
with $p$ given by (\ref{spusd2x}). Similarly, we shall denote with $p_{\rm v}(k)$ the probability that Eve sends to 
Bob $k+2$ vacuum states. This probability is given by
\begin{equation}\label{pv_k}
p_{\rm v}(k) = \left\{ \begin{array}{ll}
p^k(1-p) & \textrm{if $0\leq{}k<M_{\rm min}$}\\
(1-q)p^{M_{\rm min}}(1-p) & \textrm{if $k=M_{\rm min}$}\\
0 & \textrm{otherwise.}
\end{array} \right.
\end{equation}
We illustrate all these possible cases in figure~\ref{strategy1_meas}, where we also include the different a priori 
probabilities to be in each of these scenarios. 

Next, we obtain an expression for the gain, {\it i.e.}, the probability that Bob obtains a click per signal 
state sent by Alice, together with the quantum bit error rate (QBER) introduced by Eve with this 
sequential attack. The analysis is analogous to that included in \cite{curty_dps3}, but now taking into 
account the a priori probabilities $p_{\rm s}(k)$ and $p_{\rm v}(k)$ given by (\ref{ps_k}) and (\ref{pv_k}), respectively.

\subsection{Gain}\label{gain_A}

The gain, that we shall denote as $G$, can be expressed as $G=N_{\rm clicks}/N_{\rm s}$, where $N_{\rm clicks}$ 
represents the average total number of clicks obtained by Bob, and $N_{\rm s}$ is the total number of signal states
sent by Alice. The parameter $N_{\rm clicks}$ can be expressed as 
$N_{\rm clicks}=(N_{\rm s}/N^{\rm e})N_{\rm clicks}^{\rm e}$, 
with $N^{\rm e}$ denoting the average total number of pulses of the signal states sent by Eve 
(see figure~\ref{strategy1_meas}), and where $N_{\rm clicks}^{\rm e}$ represents the average total number of clicks
obtained by Bob when Eve sends to him precisely these signal states. With this notation, the gain of a sequential 
attack can be written as
\begin{equation}\label{gain}
G=\frac{N_{\rm clicks}^{\rm e}}{N^{\rm e}}.
\end{equation}

Let us start by calculating $N_{\rm clicks}^{\rm e}$. Whenever Eve sends to Bob a signal state $\ket{\psi_{\rm e}^k}$ 
followed by $1+d$ vacuum states (Case A in figure~\ref{strategy1_meas}) Bob always obtains one click
in his detection apparatus. On the other hand, if Eve sends to Bob only vacuum states (Case B in
figure~\ref{strategy1_meas}) Bob never obtains a click. This means, in particular, that $N_{\rm clicks}^{\rm e}$ can 
be expressed as
\begin{equation}\label{bv}
N_{\rm clicks}^{\rm e}=\sum_{k=M_{\rm min}}^{M_{\rm max}} p_{\rm s}(k)=p^{M_{\rm min}}(p+q-pq).
\end{equation}
The analysis to obtain $N^{\rm e}$ is similar. A signal state $\ket{\psi_{\rm e}^k}$ followed by $1+d$ vacuum 
states can be seen as containing $k+2+d$ pulses. On the other hand, the number of vacuum
pulses alone that Eve sends to Bob can vary from $2$ to $M_{\rm min}+2$ (see figure~\ref{strategy1_meas}).
Adding all these terms together, and taking into account their a priori probabilities, we obtain that $N^{\rm e}$ 
can be written as
\begin{eqnarray}
{}N^{\rm e}&=&\sum_{k=0}^{M_{\rm max}} p_{\rm v}(k)(k+2)+p_{\rm s}(k)(k+2+d)\nonumber \\
&=&\frac{2-p-p^{M_{\rm max}+1}}{1-p}+dN_{\rm clicks}^{\rm e},
\end{eqnarray}
with $N_{\rm clicks}^{\rm e}$ given by (\ref{bv}).

The gain $G$ can be related with a transmission distance $l$ for a given QKD scheme, {\it i.e.}, a distance which 
provides an expected click rate at Bob's side given by $G$. This last condition can be written as 
\begin{equation}\label{nuevag}
G=1-\exp{(-\mu_\alpha\eta_{\rm det}\eta_{\rm t})},
\end{equation}
where $\eta_{\rm det}$ represents the detection efficiency of Bob's detectors, and $\eta_{\rm t}$ denotes the transmittivity 
of the quantum channel. In the case of a DPS QKD scheme, the value of $\eta_{\rm t}$ can be derived from the loss 
coefficient $\gamma$ of the optical fiber measured in dB/km, the transmission distance $l$ measured in km, and the 
loss in Bob's interferometer $L$ measured in dB as
\begin{equation}\label{nuevag2}
\eta_{\rm t}=10^{-\frac{\gamma{}l+L}{10}}.
\end{equation}
From (\ref{nuevag}) and (\ref{nuevag2}), we find that the transmission distance 
$l$ that provides a gain $G$ is given by
\begin{equation}
l=-\frac{1}{\gamma}\bigg[L+10{\log_{10}}\bigg(\frac{-\ln{(1-G)}}{\mu_\alpha\eta_{\rm det}}\bigg)\bigg].
\end{equation} 

\subsection{Quantum bit error rate}\label{qber_A}

The QBER, that we shall denote as $Q$, is defined as $Q=N_{\rm errors}/N_{\rm clicks}$, where $N_{\rm errors}$ 
represents the average total number of errors obtained by Bob, and $N_{\rm clicks}$ is again the average total 
number of clicks at Bob's side. The parameter $N_{\rm errors}$ can be expressed as 
$N_{\rm errors}=(N_{\rm s}/N^{\rm e})N_{\rm errors}^{\rm e}$, with $N_{\rm errors}^{\rm e}$ denoting the average total number 
of errors obtained by Bob when Eve sends him the different signal states considered in her strategy (see
figure~\ref{strategy1_meas}). With this notation, and using again the fact that 
$N_{\rm clicks}=(N_{\rm s}/N^{\rm e})N_{\rm clicks}^{\rm e}$, we obtain that the QBER of a sequential attack can be 
expressed as
\begin{equation}\label{eqqber_g}
Q=\frac{N_{\rm errors}^{\rm e}}{N_{\rm clicks}^{\rm e}}.
\end{equation}
The parameter $N_{\rm clicks}^{\rm e}$ was calculated in the previous section and it is given by (\ref{bv}). 
In order to obtain an expression for $N_{\rm errors}^{\rm e}$, one can distinguish the same cases like in the previous 
section, depending on the type of signal states that Eve sends to Bob. Whenever Eve sends to Bob a signal state 
$\ket{\psi_{\rm e}^k}$ followed by $1+d$ vacuum states (Case A in figure~\ref{strategy1_meas}),
the average total number of errors in this scenario, that we shall denote as $e(k)$, is given by 
\begin{equation}\label{eq_ek}
e(k)=\frac{1}{2}\Bigg(1-\sum_{n=1}^{k}
\vert{}A_{n+1}^{(k)}A_n^{(k)}\vert{}\Bigg).
\end{equation}
On the other hand, if Eve sends to Bob only vacuum states (Case B in figure~\ref{strategy1_meas}) Bob 
never obtains an error. This means, in particular, that $N_{\rm errors}^{\rm e}$ can be expressed as
\begin{equation}
N_{\rm errors}^{\rm e}=\sum_{k=M_{\rm min}}^{M_{\rm max}} p_{\rm s}(k)e(k).
\end{equation}

\subsection{Evaluation}\label{primera_evalua}

The sequential attack introduced in section \ref{attack} can be parametrized by the minimum number $M_{\rm min}$ 
of consecutive unambiguous discrimination successful results that Eve needs to obtain in order to consider the 
sequence of measurement outcomes successful, the maximum number $M_{\rm max}$ of consecutive successful 
results that Eve can obtain in order to send to Bob a train of signal states, the value of the probability 
$q$, {\it i.e.}, the probability that Eve actually decides to send to Bob the signal state $\ket{\psi_{\rm e}^{M_{\rm min}}}$ 
followed by $1+d$ vacuum states instead of $M_{\rm min}+2$ vacuum states, and the state coefficients
$A_n^{(k)}\in{}\mathbb{C}$ that characterize the signal states $\ket{\psi_{\rm e}^{k}}$, with 
$M_{\rm min}\leq{}k\leq{}M_{\rm max}$.

\begin{figure}
\begin{center}
\includegraphics[scale=0.45]{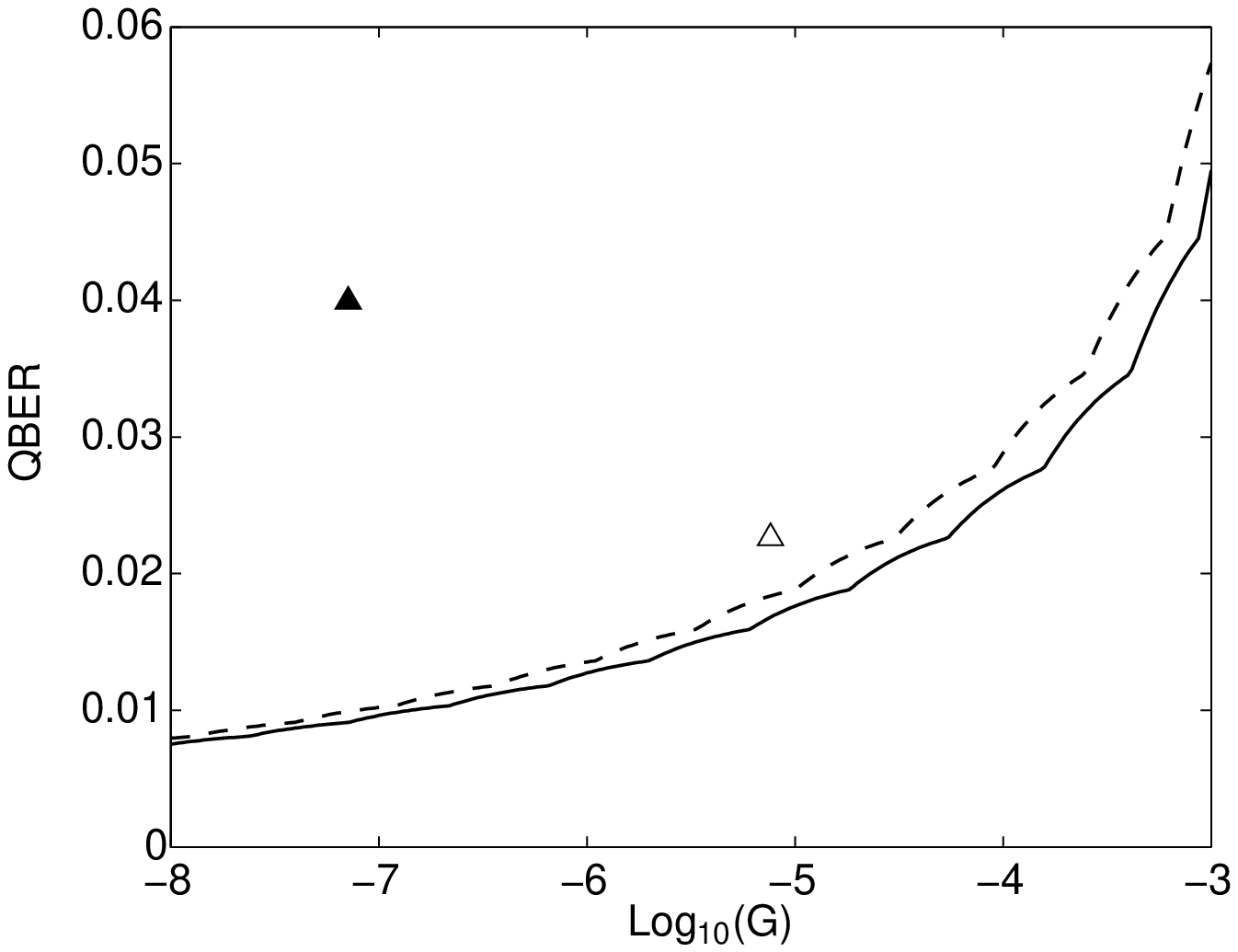}
\end{center}
\caption{Gain ($G$) versus QBER in the sequential attack introduced in section \ref{attack} for the optimal distribution 
of the state coefficients $A_n^{(k)}$ (solid line). The dashed line represents a sequential USD attack \cite{curty_dps3}. 
The mean photon number of Alice's signal states is $\mu_{\alpha}=0.2$, and the parameter $d=500$. 
The triangles represent experimental data from \cite{dpsqkd_exp3}. \label{exp1011}}
\end{figure}

\begin{figure}
\begin{center}
\includegraphics[scale=0.45]{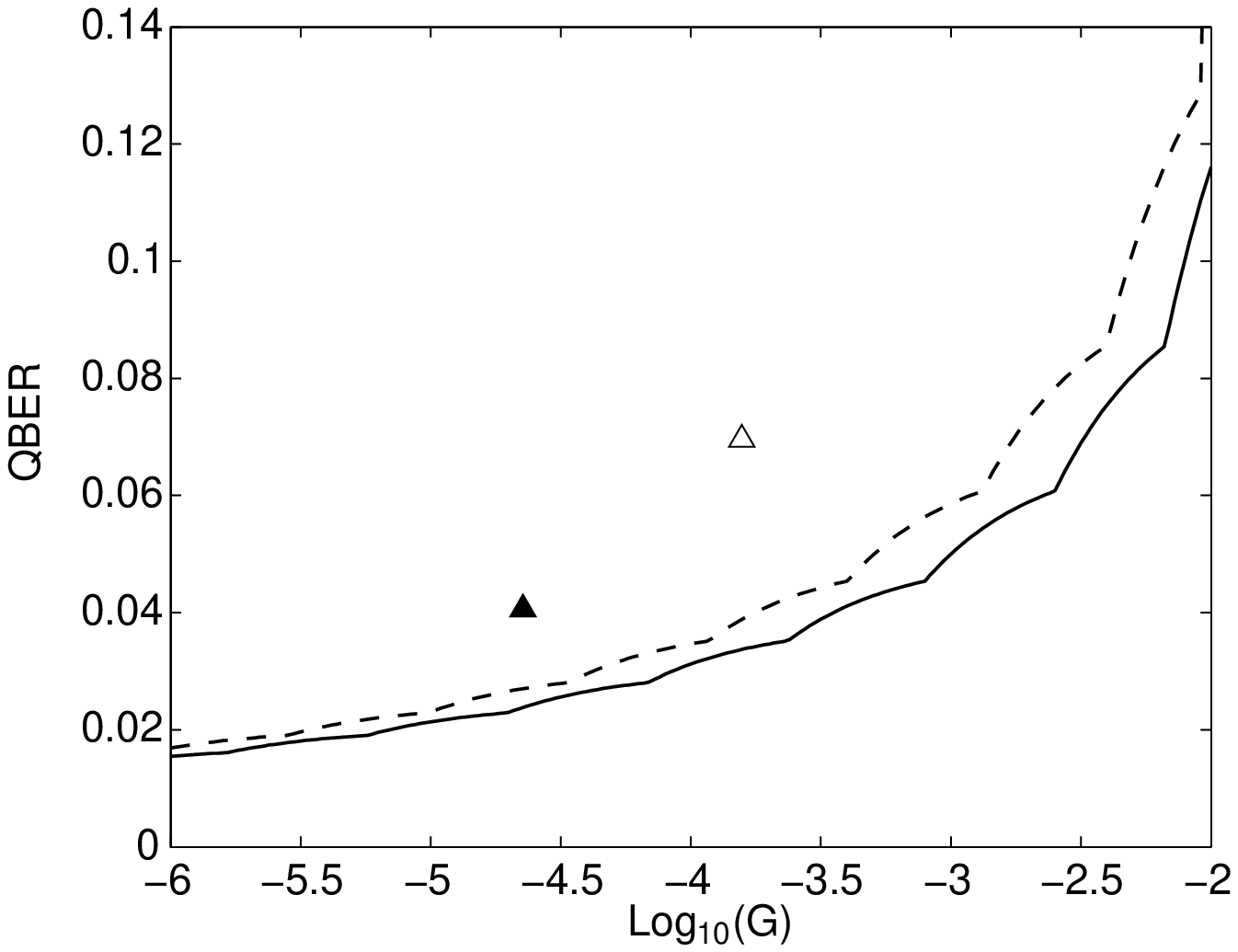}
\end{center}
\caption{Gain ($G$) versus QBER in the sequential attack introduced in section \ref{attack} for the optimal distribution 
of the state coefficients $A_n^{(k)}$ (solid line). The dashed line represents a sequential USD attack \cite{curty_dps3}. 
The mean photon number of Alice's signal states is $\mu_{\alpha}=0.17$, and the parameter $d=50$. 
The triangles represent experimental data from \cite{dpsqkd_exp1}. (See also \cite{dpsqkd_exp2b}.) \label{exp34}}
\end{figure}

\begin{figure}
\begin{center}
\includegraphics[scale=0.45]{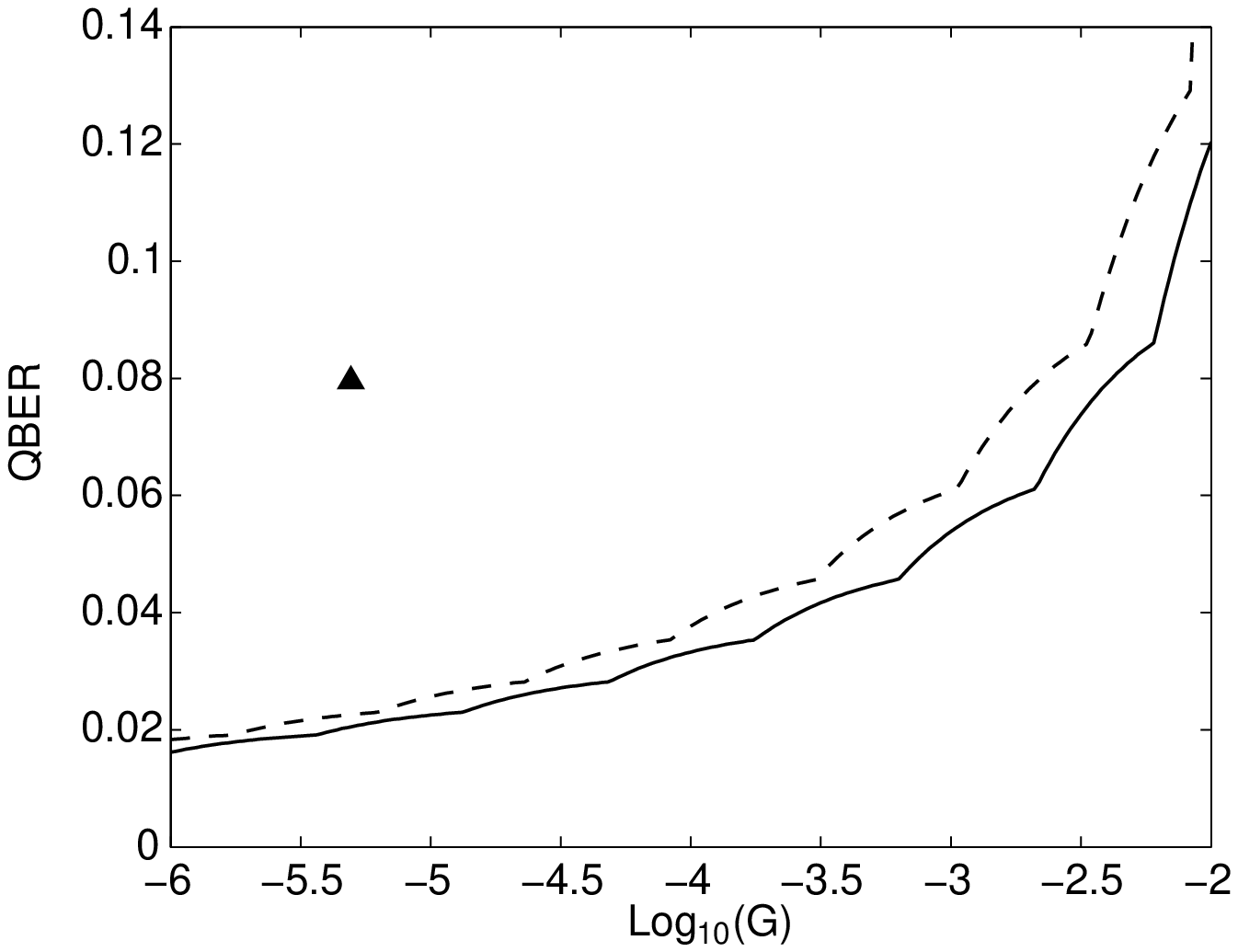}
\end{center}
\caption{Gain ($G$) versus QBER in the sequential attack introduced in section \ref{attack} for the optimal distribution 
of the state coefficients $A_n^{(k)}$ (solid line). The dashed line represents a sequential USD attack \cite{curty_dps3}. 
The mean photon number of Alice's signal states is $\mu_{\alpha}=0.16$, and the parameter $d=50$. 
The triangle represents experimental data from \cite{dpsqkd_exp1}. (See also \cite{dpsqkd_exp2b}.) \label{exp5}}
\end{figure}

\begin{figure}
\begin{center}
\includegraphics[scale=0.45]{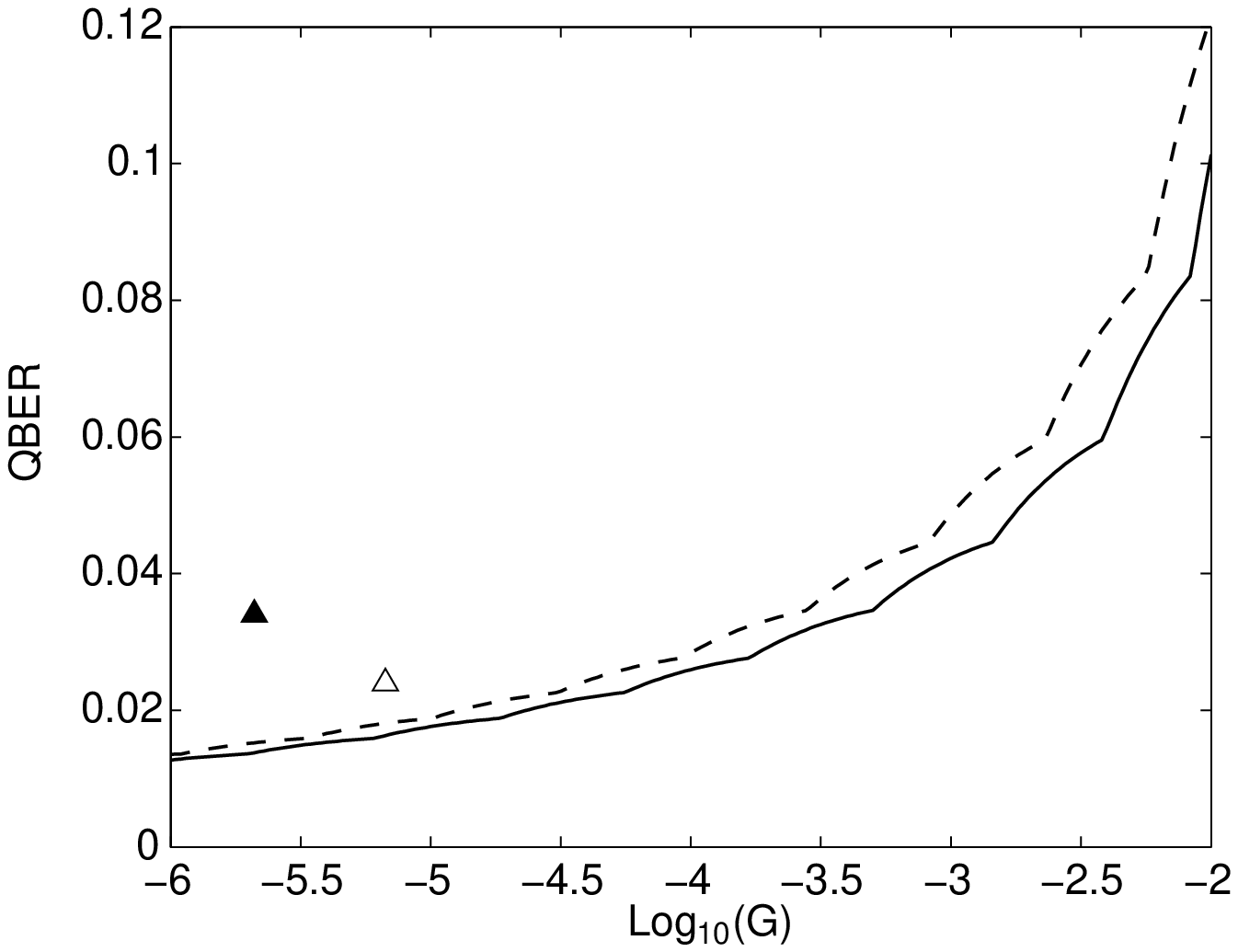}
\end{center}
\caption{Gain ($G$) versus QBER in the sequential attack introduced in section \ref{attack} for the optimal distribution 
of the state coefficients $A_n^{(k)}$ (solid line). The dashed line represents a sequential USD attack \cite{curty_dps3}. 
The mean photon number of Alice's signal states is $\mu_{\alpha}=0.2$, and the parameter $d=50$. 
The triangles represent experimental data from \cite{dpsqkd_exp2}. (See also \cite{dpsqkd_exp2b}.) \label{exp69}}
\end{figure}

Figures~\ref{exp1011}, \ref{exp34}, \ref{exp5} and \ref{exp69} show a graphical representation of the gain
versus the QBER in this sequential attack for different values of the mean photon number $\mu_{\alpha}$ of Alice's signal 
states, and the parameter $d$. It states that no key distillation protocol can provide a secret key from the correlations established by the users above the curves, {\it i.e.}, the secret key rate in that region is zero. In these examples we 
consider the optimal distribution for the state coefficients $A_n^{(k)}$, {\it i.e.}, the one which provides the lowest 
QBER for a given value of the gain. This distribution was obtained in \cite{curty_dps3}, where it was shown that 
the vector of optimal state coefficients $(A_1^{(k)}, ..., A_{k+1}^{(k)})$ coincides with the normalized 
eigenvector associated with the maximal eigenvalue of a 
$(k+1) \times (k+1)$ matrix with ones only on the first off-diagonals and zeros elsewhere. 
These figures assume that $M_{max}$ is fixed 
and given by $M_{max}=25$, and we vary the parameters $M_{min}<M_{max}$ and $q\in[0,1]$.  
These examples also include the case of a sequential attack where
Eve realizes USD of each signal state sent by Alice \cite{curty_dps3}, together with experimental data from 
\cite{dpsqkd_exp1,dpsqkd_exp2,dpsqkd_exp2b,dpsqkd_exp3}. For instance, in the experiment reported in \cite{dpsqkd_exp3} the dead-time of Bob's detectors is $t_{\rm d}=50$ ns and the clock frequency of the system is 
$f_{\rm c}=10$ GHz. We obtain, therefore, that $d=\lceil{}t_{\rm d}f_{\rm c}\rceil=500$. (See figure~\ref{exp1011}.) 
Similarly, in the experiments realized in \cite{dpsqkd_exp1,dpsqkd_exp2,dpsqkd_exp2b} we have that $t_{\rm d}=50$ 
ns and $f_{\rm c}=1$ GHz. This means, in particular, that in all these cases $d=50$. (See figures~\ref{exp34}, 
\ref{exp5} and \ref{exp69}.)

According to these results, we find that the sequential attack proposed in section \ref{attack} can provide tighter upper 
bounds for 
the security of a DPS QKD scheme than those derived from a sequential attack where Eve performs USD of each 
signal state emitted by the source. Basically, this result arises due to the different 
a priori probabilities of Eve sending to Bob a train of signal states $\ket{ \psi_{\rm e}^{k}}$, together with $1+d$ vacuum 
states, in each of these two possible attacks. In particular, while in the attack introduced in section \ref{attack}
these probabilities are given by $p_{\rm s}(k)$, in a sequential 
USD attack these probabilities have the form $p_{\rm s}(k)p$, with $p$ given by (\ref{spusd2x}). Note that in this last 
case Eve has to discriminate the state of $k+1$ consecutive signals sent by Alice unambiguously. 
 
\section{CONCLUSION}\label{CONC}

In this paper we have analyzed limitations imposed by sequential attacks on the performance of a 
differential-phase-shift (DPS) quantum key distribution (QKD) protocol based on weak coherent pulses. 
A sequential attack consists of Eve measuring out every coherent state emitted by Alice and, afterwards, 
she prepares new signal states, depending on the results obtained, that are given to Bob. Whenever 
Eve obtains a predetermined number of consecutive successful measurement outcomes, 
then she prepares a new train of non-vacuum signal states that is forwarded to Bob. Otherwise,
Eve can send vacuum signals to Bob to avoid errors. Sequential attacks transform the original quantum 
channel between Alice and Bob into an entanglement breaking channel and, therefore, they do not allow 
the distribution of quantum correlations needed to establish a secret key. 

Specifically, we have investigated a sequential attack where Eve realizes optimal unambiguous discrimination 
of the relative phases between Alice's signal states. When Eve identifies unambiguously the relative 
phase between two consecutive signal states sent by Alice, then she considers this result as successful. Otherwise, 
she considers it a failure. As a result, we obtained ultimate upper bounds on the maximal distance 
achievable by a DPS QKD scheme as a function of the error rate in the sifted key, and the mean photon number 
of Alice's signals. It states that there exists no improved classical
communication protocol or improved security analysis which can turn the correlations 
established by the users into a secret key. Moreover, our analysis indicates that this attack can provide 
tighter upper bounds for the security of a DPS QKD scheme than those derived 
from sequential attacks where Eve performs unambiguous state discrimination of each signal state emitted by the 
source.  

\section{ACKNOWLEDGEMENTS}

The authors wish to thank Norbert L\"utkenhaus and Tobias Moroder for very fruitful discussions on the topic of this 
paper and very useful comments on the manuscript. M.C. especially thanks Norbert L\"utkenhaus for hospitality and 
support during his stay at the Institute for Quantum Computing (University of Waterloo) where this manuscript was 
finished.
  
\appendix

\section{Optimality of Eve's measurement}\label{ap_opt}

In this appendix we show that the unambiguous discrimination measurement presented in section \ref{meas}
is optimal, {\it i.e.}, it minimizes the probability of having an inconclusive result when distinguishing all the relative 
phases between Alice's signal states. 
For that, we calculate the maximal probability of unambiguously determining all the relative phases contained in 
the signal states $\rho{}(\vec x_M)$ given by (\ref{yns}), and we show that this probability coincides with that 
provided by the measurement introduced in section \ref{meas}. 

As already mentioned before, due to the special block structure of the signal states $\rho{}(\vec x_M)$, 
we can always assume, without loss of generality, that Eve first projects $\rho{}(\vec x_M)$
onto the orthogonal subspaces $\mathcal{A_M}$ and  $\mathcal{B_M}$ and, afterwards, she measures 
the relative phase information contained in $\ket{\psi_{y}(\vec x_M)}$, with $y\in\{A,B\}$. 

The set of states $\ket{\psi_{y}(\vec x_M)}\in\mathcal{Y_M}$ constitutes a so-called geometrically uniform (GU) set
\cite{for91,eldar}. That is, these states are defined over a group of unitary matrices 
and they can be generated by a single generating vector. In particular, let $\mathcal{G}$ be the finite group of 
$2^{M-1}$ unitary matrices $U(\vec x_M)$ defined as
\begin{eqnarray}
U(\vec x_M)&=&\sum_{n_1, ..., n_M=0}^{1} (-1)^{\sum_{i=1}^{M-1}x_in_i} \ket{n_1, ..., n_M}\nonumber \\
&\times&\bra{n_1, ..., n_M},
\end{eqnarray}
with $\vec x_M$ given by (\ref{vecx}). If we denote as $\vec 0_M=(0_1, ..., 0_M)$ the vector that has all its $M$ 
elements equal to zero, then the states $\ket{\psi_{y}(\vec x_M)}$ can always be written as 
$\ket{\psi_{y}(\vec x_M)}=U(\vec x_M)\ket{\psi_{y}(\vec 0_M)}$, with $\ket{\psi_{y}(\vec 0_M)}$ being the 
generating vector of the set. 

Let $\Phi_{y,M}$ denote the matrix whose columns are the state vectors $\ket{\psi_{y}(\vec x_M)}$, and let 
$\Phi_{y,M}^*$ represent its conjugate transpose. It was proven in \cite{eldar} that the maximal probability of 
correctly distinguishing between GU pure states with equal a priori probabilities
is given by the smallest eigenvalue of $\Phi_{y,M}\Phi_{y,M}^*$. The matrices $\Phi_{y,M}$, with $y\in\{A,B\}$ and 
$M\geq{}3$, can be written, respectively, as
\begin{widetext}
\begin{eqnarray} 
\Phi_{A,M} &=& 
\frac{1}{\sqrt{p_{A,M}}}
\left( \begin{array}{cc} 
a\sqrt{p_{A,M-1}}\Phi_{A,M-1} & a\sqrt{p_{A,M-1}}\Phi_{A,M-1} \\ 
b\sqrt{p_{B,M-1}}\Phi_{B,M-1} & -b\sqrt{p_{B,M-1}}\Phi_{B,M-1}  
\end{array} \right), \nonumber \\
\Phi_{B,M} &=& \frac{1}{\sqrt{p_{B,M}}}
\left( \begin{array}{cc} 
a\sqrt{p_{B,M-1}}\Phi_{B,M-1} & -a\sqrt{p_{B,M-1}}\Phi_{B,M-1} \\ 
b\sqrt{p_{A,M-1}}\Phi_{A,M-1} & b\sqrt{p_{A,M-1}}\Phi_{A,M-1}  
\end{array} \right),
\end{eqnarray} 
\end{widetext}
where $\Phi_{y,M-1}$ denotes the matrix whose columns are the state vectors $\ket{\phi_{y}(\vec x_{M-1})}$
given by (\ref{states_tilde}). This means, in particular, that $\Phi_{y,M}\Phi_{y,M}^*$ can be expressed as 
a block-diagonal matrix as
\begin{widetext}
\begin{equation} 
\Phi_{y,M}\Phi_{y,M}^* = \frac{1}{p_{y,M}}
\left( \begin{array}{cc} 
2a^2p_{y,M-1}\Phi_{y,M-1}\Phi_{y,M-1}^* & \bar 0 \\ 
\bar 0 & 2b^2p_{{\bar y},M-1}\Phi_{{\bar y},M-1}\Phi_{{\bar y},M-1}^*  
\end{array} \right),
\end{equation} 
\end{widetext}
with $\bar 0$ denoting a zero matrix, {\it i.e.}, a matrix which contains only zeros. 
The smallest eigenvalue of $\Phi_{y,M}\Phi_{y,M}^*$, that we shall denote as $\lambda_{y,M}^{min}$, is given by
\begin{equation}\label{eig} 
\lambda_{y,M}^{min}=
\textrm{min }\Bigg\{\frac{2a^2p_{y,M-1}\lambda_{y,M-1}^{min}}{p_{y,M}},
\frac{2b^2p_{{\bar y},M-1}\lambda_{{\bar y},M-1}^{min}}{p_{y,M}}\Bigg\},
\end{equation}
with $\lambda_{y,M-1}^{min}$ denoting the smallest eigenvalue of $\Phi_{y,M-1}\Phi_{y,M-1}^*$. We solve (\ref{eig}) by 
induction. In particular, we start by analyzing the case $M=2$, and then we show that 
\begin{equation}\label{compara}
2a^2p_{y,M-1}\lambda_{y,M-1}^{min}\geq{}2b^2p_{{\bar y},M-1}\lambda_{{\bar y},M-1}^{min}, 
\end{equation}
 for all $M\geq{}3$.
 
When $M=2$, we have that $\Phi_{A,2}\Phi_{A,2}^*=(p_{A,2})^{-1}[2a^4, 0; 0, 2b^4]$, and 
$\Phi_{B,2}\Phi_{B,2}^*=(p_{B,2})^{-1}[2a^2b^2, 0; 0, 2a^2b^2]$. Then, since $a>b$, it is guaranteed that
$2a^4=2a^2p_{A,1}>2b^4=2b^2p_{B,1}$, and $2a^2b^2=2a^2p_{B,1}=2b^2p_{A,1}$,
respectively. That is, if we define $\lambda_{y,1}^{min}=1$ for all $y\in\{A,B\}$, then (\ref{compara}) is satisfied.
When $M=3$, it turns out that $2a^2p_{A,2}\lambda_{A,2}^{min}=
2b^2p_{B,2}\lambda_{B,2}^{min}=4a^2b^4$, and 
$2a^2p_{B,2}\lambda_{B,2}^{min}=4a^4b^2>
2b^2p_{A,2}\lambda_{A,2}^{min}=4b^6$. That is, (\ref{compara}) is also satisfied. Let us now assume 
that 
$2a^2p_{y,M-2}\lambda_{y,M-2}^{min}\geq{}2b^2p_{{\bar y},M-2}\lambda_{{\bar y},M-2}^{min}$ is true. 
Then, from (\ref{eig}) we have that 
$2a^2p_{y,M-1}\lambda_{y,M-1}^{min}=2a^2 \textrm{min }\{2a^2p_{y,M-2}\lambda_{y,M-2}^{min},
2b^2p_{{\bar y},M-2}\lambda_{{\bar y},M-2}^{min}\}
=2a^22b^2p_{{\bar y},M-2}\lambda_{{\bar y},M-2}^{min}
\geq{} 2b^22b^2p_{y,M-2}\lambda_{y,M-2}^{min}=2b^2 p_{{\bar y},M-1}\lambda_{{\bar y},M-1}^{min}$. 

This means, 
therefore, that $\lambda_{y,M}^{min}$ is given by
\begin{eqnarray}\label{cet}
\lambda_{y,M}^{min}&=&\frac{1}{p_{y,M}}\bigg(2b^2p_{{\bar y},M-1}
\lambda_{{\bar y},M-1}^{min}\bigg)\nonumber \\
&=&p_{succ,y,M}\lambda_{{\bar y},M-1}^{min},
\end{eqnarray}
where in the last equality we have used (\ref{lms}). When $M$ is even, this expression can be written as
\begin{equation}\label{chi}
\lambda_{y,M}^{min}=\prod_{i=1}^{M/2}p_{succ,y,2i}\prod_{j=1}^{M/2-1}p_{succ,{\bar y},2j+1}, 
\end{equation}
while, whenever $M$ is odd then (\ref{cet}) has the form
\begin{equation}\label{ki}
\lambda_{y,M}^{min}=\prod_{i=1}^{(M-1)/2}p_{succ,y,2i+1}p_{succ,{\bar y},2i}.
\end{equation}

The maximal probability of unambiguously determining all the relative phases contained in 
the signal states $\rho{}(\vec x_M)$ is given by
\begin{equation}
\sum_{y\in\{A,B\}} p_{y,M} \lambda_{y,M}^{min}.
\end{equation}
After some straightforward calculations,  we obtain that this quantity can be written as 
$[1-\exp{(-2\mu_{\alpha})}]^{M-1}$, which coincides with that obtained in section \ref{meas}.

\bibliographystyle{apsrev}
\bibliographystyle{apsrev}

\begin{thebibliography}{99}
\bibitem{note} In order to simplify our notation, 
from now on we will employ the term ``signal state" only to denote those light pulses with a mean photon 
number bigger than zero. A light pulse with an average photon number equal to zero will be always denoted
as a ``vacuum state".
\end{thebibliography}

\end{document}